\documentclass[conference]{IEEEtran}

\ifCLASSINFOpdf
\else
\fi

\usepackage{algorithm}
\usepackage{algorithmic}
\usepackage{multirow}
\usepackage{graphicx}
\usepackage{epstopdf}

\usepackage{amsmath}
\usepackage{amssymb}
\usepackage{subfigure}
\usepackage{url}
\usepackage{epstopdf}
\usepackage{amssymb}
\usepackage{booktabs}
\usepackage{mathtools}
\usepackage{enumerate}
\usepackage{url}
\usepackage{nomencl}
\usepackage{color}
\begin{document}
%
\title{Routing Diverse Evacuees with Cognitive Packets}


\author{Huibo Bi and Erol Gelenbe\\
Imperial College London\\
Department of Electrical and Electronic Engineering\\
Intelligent Systems and Networks Group\\
Email: \{huibo.bi12, e.gelenbe\}@imperial.ac.uk}
\maketitle

\begin{abstract}
This paper explores the idea of smart building evacuation when evacuees can belong to different
categories with respect to their ability to move and their health conditions. This leads to new algorithms
that use the Cognitive Packet Network concept to tailor different quality of service needs to different evacuees.
These ideas are implemented in a simulated environment and evaluated with regard to their effectiveness.
\end{abstract}


%
\IEEEpeerreviewmaketitle

\section{Introduction}

Emergency navigation \cite{DBES,CAMWA,Wu2} is the process of guiding evacuees to exits when an
emergency occurs, and this paper presents an multipath routing algorithm that attempts to satisfy the requirements of diverse classes of evacuees. The approach is based on the Cognitive Packet Network (CPN) which searches \cite{Search,Cao,PhysRev2010} for optimised routes rapidly and adaptively with regard to user-defined Quality of Service (QoS) goals to evaluate the quality of paths and choose appropriate routes for different groups of evacuees. In order to minimise the congestion encountered by each evacuee, a congestion-aware algorithm that can predict the future clogging with respect to the current observed congestion is proposed. ``Direction oscillation problem'' \cite{chen2008load} that occurs in adaptive routing with sensitive metrics \cite{Sensible} due to delays in the available information is discussed and ``movement depth'' is introduced to facilitate the evacuees to update their routes periodically to avoid being directed into hazards.

The remainder of the paper is organized as follows: Section \ref{Related} presents the related studies of emergency navigation algorithms. In Section \ref{CPN} we recall the concept of CPN. Then we introduce the QoS metrics designed for diverse categories of evacuees in Section \ref{QoS}, and discuss the issue of oscillations that arise in adaptive networks and their alleviation in Section \ref{Oscill}. The simulation model and assumptions are introduced in Section \ref{Sim}. Finally, the results of experiments are discussed in Section \ref{Discuss}.

\section{Related Work} \label{Related}

Much research in emergency navigation focuses on ``normal'' individuals with identical physical attributes such as mobility and health level, and conventional models such as flow-based models \cite{chalmet1982network,francis1984negative,kisko1985evacnet+,hoppe1994polynomial,hoppe2000quickest,lu2003evacuation} hypothesise evacuees as continuous homogeneous flows. Potential-maintenance approaches \cite{li2003distributed,tseng2006wireless,park1997highly,pan2006emergency} concentrate on the navigation algorithms that do not take the physical attributes of evacuees into account. The ``magnetic model'' in \cite{okazaki1993study} sets a walking velocity for each evacuee but ignores personal requirements as well as the effect of social interaction. The advent of multi-agent models makes it more convenient to customise physical attributes for individuals. However, previous work has focused on incorporating sociological factors due to the easiness of describing social behaviours such as coordination or stampede. Although in \cite{bo2009multi} physical, psychological and moving attributes are considered for each evacuee, health factors such as initial health values and resistance to hazard which are influenced by gender, age, etc. are not considered. Queueing models can also be used to evaluate congestion in such systems \cite{ActaI}\cite{ActaII}.

Navigation algorithms proposed previously search for the shortest or the safest path and in \cite{li2003distributed} a self-organizing sensor network is proposed to guide a user (robots, people, unmanned aerial vehicle, etc.) through the safest path by using ``artificial potential fields'' \cite{Search}. An attractive force pulls users to the destination while repulsive forces from dangerous zones push them away. In \cite{tseng2006wireless} a temporally ordered routing algorithm \cite{park1997highly} routes evacuees to exits through safer paths. A navigation map is manually defined to avoid impractical paths and each sensor is assigned with an altitude with respect to its hops to the nearest exit. Combining the definition of effective length with Dijkstra's algorithm, \cite{filippoupolitis2009distributed} presents a decentralized evacuation system with decision nodes (DNs) and sensor nodes (SNs) to compute the shortest routes in real time, while opportunistic communications \cite{pelusi2006opportunistic} based emergency evacuation systems \cite{Procedia}
have the advantage of being more robust to network attacks \cite{DoS} that often accompany emergencies caused by malicious acts.

\section{The Cognitive Packet Network} \label{CPN}

The Cognitive Packet Network (CPN) introduced in \cite{gelenbe2001towards,gelenbe2004self,gelenbe2009steps} was originally proposed for  large scale and fast-changing networks. As the conditions of networks typically change frequently, the protocols which periodically update the network status as well as the routing schemes that require the convergence of an entire network can always result in a constant lag and impede the performance. Therefore, the concept of CPN is proposed to solve this problem by sending Cognitive Packets to monitor the network conditions and discover new routes sequentially. In the CPN, ``Cognitive packets'' play a dominating role in routing and flow control other than routers and protocols. Cognitive packets can adapt to pursue their predefined goals and learn from their own investigations and experience from other packets.
Each node in CPN hosts recurrent random neural networks (RNN) with a neuron per neighbour node, and maintains a routing list which reserves a fixed number of QoS-oriented routes in a descending order. CPN carries three types of packets: smart packets (SPs), acknowledgements (ACKs) and dumb packets (DPs). SPs are used for information collection and routing discovery with respect to a specific QoS goal. They can either choose the most excited neuron as next hop or drift randomly to explore and measure undiscovered or not recently measured paths. When a SP reaches the destination, an ACK which stores all the gathered information will be generated and sent back to the source node through the reverse path. Because loops may be produced during route exploration, a loop remove algorithm is acted on the reverse path to take out any sequences of nodes with the same origin and destination. When an ACK arrives at a node, it will update the routing list and train a random neural network \cite{RNN2008} by performing Reinforcement Learning (RL)
with neural networks \cite{gelenbe2001simulations,Natural}. DPs are the packets which actually carry the payloads. It always selects the best path at the top of the routing list as the next hop.

In the context of emergency evacuation, because the environment is highly dynamic, SPs
are used to search available routes and collect information such as the hazard intensity of a
node. ACKs backtrack with collected information and update the excitation level of neurons at each traversed node. Evacuees are considered as DPs and always follow the top ranked path in the routing list. To apply the concept of CPN to emergency evacuation, the following requirements
must be satisfied: a predefined graph-based layout of a confined place with information such
as length of edges and node capacity is available; The edges of the graph depict the paths and the vertices represent the physical areas where sensors are installed. The sensors could communicate with its neighbour nodes and detect the typical hazards such as fire, smoke,
water, etc. Additionally, in order to perform the congestion-aware algorithm, we also assume that the sensors can sense the arrival of a civilian and record the number of evacuees congested in proximity.

\section{QoS Metrics} \label{QoS}
QoS metrics are expressed as goal functions which provide input data for RNN and are minimized by CPN. Details of the RNN based algorithm are available in \cite{gelenbe1999cognitive}. It is not included in this paper due to space constraints. The simplest metric used in the CPN based algorithm in \cite{bib:BiDesmetGelenbeISCIS2013} is  distance based, and it pursues the shortest effective path, rather than the path with the smallest number of segments or hops to traverse, from a source to a destination. Other novel metrics considered in this paper are the time-oriented goal function, the energy efficiency oriented goal function and the safety-oriented goal function. However our simulations will be limited to the time oriented goal. Previous work in \cite{desmet2013graph} has shown that congestion will affect the performance of evacuation routing algorithms significantly. Meanwhile, certain strategic nodes such as stairs can become bottlenecks and cause long lasting congestion during an evacuation process. In order to ease congestion and bypass bottleneck nodes appropriately, we include congestion in the QoS metric.

Unlike previous congestion-aware algorithms, the algorithm that we proposed evaluates if the current congestion located at a certain node will affect an evacuee by comparing the duration of the congestion and the arrival instant of the evacuee. The congestion situation at a node when an evacuee reaches depends on the current situation of the node and the total number of arrivals and departures before the evacuee arrives. To estimate the total number of arrivals and departures, we use the average arrival rate and departure rate recorded by each node. The details are given in \textbf{Algorithm 1} and we assume that if the queue length of a node is larger than one, then congestion will occur.
\begin{algorithm}[!htb]
\caption{Predict the potential amount of congestion encountered by an evacuee during the evacuation process}
\textbf{Data:}\ A path $\pi$ explored by SPs from a node $\pi(0)$\\
\textbf{Result:}\ The potential number of congestion encountered by an evacuee traversing across the path $\pi$\\
\begin{algorithmic}[1]
\STATE Set the total number of congestion $C_{total}$ to 0
\STATE Set the total travel time $T_{total}$ to 0
\FORALL {the edges $e(\pi(i),\pi(i+1))$ $\in$ path $\pi$}
\STATE /* Calculate the time cost on the edge $e(\pi(i),\pi(i+1))$ */
\STATE $t_{edge} \leftarrow \frac{E^{e}(\pi(i),\pi(i+1))}{V_{speed}}$ \\ where $E^{e}(\pi(i),\pi(i+1))$ is the effective length of $e(\pi(i),\pi(i+1))$, $\pi(i)$ is the source node of the edge, $\pi(i+1)$ is the destination node of the edge, $V_{speed}$ is the average speed of certain type of evacuees.
\STATE /* Compute the time cost at the node $\pi(i)$ */
\IF {$ q_{0}^{i} + \bar{\lambda} \cdot T_{\pi(i)} - \bar{\mu} \cdot T_{\pi(i)} > 0$}
\STATE /* If the congestion will remain there when the evacuee reaches the node $\pi(i)$, then plus the time delay caused by congestion */
\STATE $t_{node} \leftarrow \frac{q_{0}^{i} + \bar{\lambda} \cdot T_{\pi(i)} - \bar{\mu} \cdot T_{\pi(i)}}{\bar{\lambda}}$ \\ where $q_{0}^{i}$ is the current queue length of the node $\pi(i)$, $\bar{\lambda}$ is the average arrival rate of the node $\pi(i)$, $\bar{\mu}$ is the departure rate of the node $\pi(i)$, $T_{\pi(i)}$ is the potential time cost to reach the node $\pi(i)$, which is the current value of $T_{total}$. Term $\frac{q_0 + \bar{\lambda} \cdot T_{\pi(i)} - \bar{\mu} \cdot T_{\pi(i)}}{\bar{\lambda}}$ is the average time calculated by Little's formula.
\STATE /* If the predicted queue length when the evacuee reaches it is larger than zero, the arrival of the evacuee will trigger congestion */
\STATE $C_{total} \leftarrow C_{total} + 1$
\ENDIF
\STATE $T_{total} \leftarrow T_{total} + t_{edge} + t_{node}$
\ENDFOR
\STATE \textbf{Return} $C_{total}$
\end{algorithmic}
\end{algorithm}

The time-oriented goal function pursues the shortest time cost to reach an exit, and satisfies the QoS need of normal evacuees. The path traversal time is predicted as follows:
\begin{equation}
\begin{split}
& G_{t} = \sum_{i=1}^{n-1}\{\frac{E^{e}(\pi(i),\pi(i+1))}{V_{speed}}\\
& + K[\frac{q^{i}_{0} + \bar{\lambda} \cdot T_{\pi(i)} - \bar{\mu} \cdot T_{\pi(i)}}{\bar{\lambda}}]\}
\end{split}
\end{equation}
where $\pi$ represents a particular path, $n$ is the number of nodes on the path $\pi$, and $\pi(i)$ is the $i$-th node on the path $\pi$. $E^{e}(\pi(i),\pi(i+1))$ is the effective length of the edge between node $\pi(i)$ and node $\pi(i+1)$. $V_{speed}$ is the average speed of certain category of civilians. $K[X]$ is a function that takes the value 0 if X is smaller than zero or X if X is larger than or equals to zero respectively. Term $q^{i}_{0}$ is the current queue length of the node $\pi(i)$, $\bar{\lambda}$ is the average arrival rate of the node $\pi(i)$, $\bar{\mu}$ is the departure rate of the node $\pi(i)$, $T_{\pi(i)}$ is the potential time cost to reach the node $\pi(i)$.

Energy efficiency oriented goals will search for the path with minimum energy consumption. This function can be used for wheelchairs or robots due to the battery power limitations.
\begin{equation}
\begin{split}
& G_{e}=c_b\cdot C_{total}+\sum_{i=1}^{n-1}c_s\cdot E^{e}(\pi(i),\pi(i+1))+ \\
& \sum_{i=2}^{n-1}c_t\cdot\theta (\pi(i-1),\pi(i),\pi(i+1))
\end{split}
\end{equation}
If $C_{total}$ is the total congestion calculated by \textbf{Algorithm 1}, then $c_b$ is the energy consumption of a wheelchair when a braking event happens, and $c_s$ is the energy consumption per centimetre when moving straight on an edge. $E^{e}(\pi(i),\pi(i+1))$ is the effective length of the edge between node $\pi(i)$ and node $\pi(i+1)$. Term $c_t$ is the energy consumption per degree when a turning event happens, $\theta (\pi(i-1),\pi(i),\pi(i+1))$ is the rotation angle between edge $E(\pi(i-1),\pi(i))$ and $E(\pi(i),\pi(i+1))$.

The safety-oriented goal function resolves the safest path. It can be used for sick people or children as they are more likely to faint due to the impact of poisonous smoke in a fire. To estimate the hazard intensity of a node, we assume the hazard spread rate is $a$ (cm/s) and the hazard growth rate at a node is $b$.
\begin{equation}
\begin{split}
& G_{s} = \sum_{i=1}^{n-1}1[t_{evacuee}^{\pi(i+1)}+t_{current}<t_{hr}^{\pi(i+1)}]\cdot b \cdot \\ & (t_{evacuee}^{\pi(i+1)}+t_{current}-t_{hr}^{\pi(i+1)})+E^{s}(\pi(i),\pi(i+1))
\end{split}
\end{equation}
where $1[X]$ is a function that takes the value 1 or 0 if X is false or true respectively. The term $t_{evacuee}^{\pi(i+1)}$ is the time for an evacuee to reach node $\pi(i+1)$ and $t_{current}$ is the current simulation time. The term $t_{hr}^{\pi(i+1)}$ is the time for the hazard to reach node $\pi(i+1)$, it can be estimated with the hazard spread rate $a$. $E^{s}(\pi(i),\pi(i+1))$ is then the effective safety value between node $\pi(i)$ and node $\pi(i+1)$. We employ $E^{s}(\pi(i),\pi(i+1))$ to ensure the value of $G_{s}$ is not nil if the hazard does not reach the path during a civilian's evacuation process.

The details of this approach is shown in \textbf{Algorithm 2}.
\begin{algorithm}
\caption{Predict the potential hazard of a path}
\textbf{Data:} A path $\pi$ explored by SPs from a node\\
\textbf{Result:} The potential hazard of the path $\pi$\\
\begin{algorithmic}[1]
\FORALL {the nodes $n$ in the environment}
\STATE Find the shortest path $D$ from $n$ to the fire source by using the built-in map
\STATE $t_{hr} \leftarrow D/a$ \\ where $t_{hr}$ is the time cost for the hazard to reach $n$
\ENDFOR
\STATE Set the total potential hazard $H_{total}$ of path $\pi$ to 0
\FORALL {the nodes $\pi(i)$ $\in$ path $P$}
\STATE /* Calculate the potential time cost $t_{evacuee}^{\pi(i)}$ for an evacuee to reach $\pi(i)$ from the source node of $\pi$ */
\STATE $t_{evacuee}^{\pi(i)}$ can be calculated by \textbf{Algorithm 1}
\STATE /* Calculate the potential hazard $h_{node}$ when an evacuee arrives $\pi(i)$ from the source node of $\pi$ */
\STATE $h_{node} \leftarrow b\cdot (t_{evacuee}^{\pi(i)}+t_{current}-t_{hr}) + E^{s}(\pi(i),\pi(i+1))$ \\ where $E^{s}(\pi(i),\pi(i+1))$ is the effective safety value of the edge between node $\pi(i)$ and $\pi(i+1)$ \\
\STATE $H_{total} \leftarrow H_{total} + h_{node}$
\ENDFOR
\STATE \textbf{Return} $H_{total}$
\end{algorithmic}
\end{algorithm}

\section{Alleviation of Oscillations} \label{Oscill}

Path oscillations will occur when a sensitive metric \cite{Sensible} is used or the process of updating sensor readings and arrival of ACKs are asynchronous when the hazard spreads quickly. This phenomenon does not affect evacuees with portable devices as they will receive a whole path but confuses civilians using visual indicators because the guiding direction oscillates. In \cite{gelenbe2009steps} several ways to mitigate oscillations in packet networks are discussed such as using a path switch probability, or a minimum QoS gain threshold to switch routes, or ensuring the usage of a path by a minimum number of packets before a path switch can occur though this last choice would not make sense if ``packets'' are human beings and the path switch is being caused by a deterioration of the safety of certain paths. In our system we set a path switch probability to ease the direction oscillation on the visual indicators.

Moveover, because CPN is a source routed algorithm, each evacuee with a portable device will obtain a route to exits when they begin to evacuate. However, some evacuees cannot stick to the original route as the path condition varies (hazard, congestion, etc.). A simple method for the evacuees to use the up-to-date suggestion is to adopt the top ranked route at each hop. However, this may cause  oscillations when successive decisions vary frequently. Hence, we use ``movement depth'' to ensure that evacuees only accept a new path choice suggestion after traversing a certain number of nodes. To prevent that movement depth hinders the flexibility of the evacuation routing and directs civilians to hazards, the sequential decisions will be checked periodically to verify if the hazard has reached a particular node. If the hazard has reached any hop or edge related to these decisions, they will be discarded immediately and an alternate path will be used.
However this approach can also be very dangerous since evacuees may get close to a hazard and then have no remaining time or possibility to change to a safe path. Thus such approaches need to be evaluated in the context of varying degrees of risk that include random rates of movement of the hazard (fire, smoke, gases, etc.) itself.

\begin{figure}[ht!]
\centering
\includegraphics[width=0.6\textwidth,height=8cm]{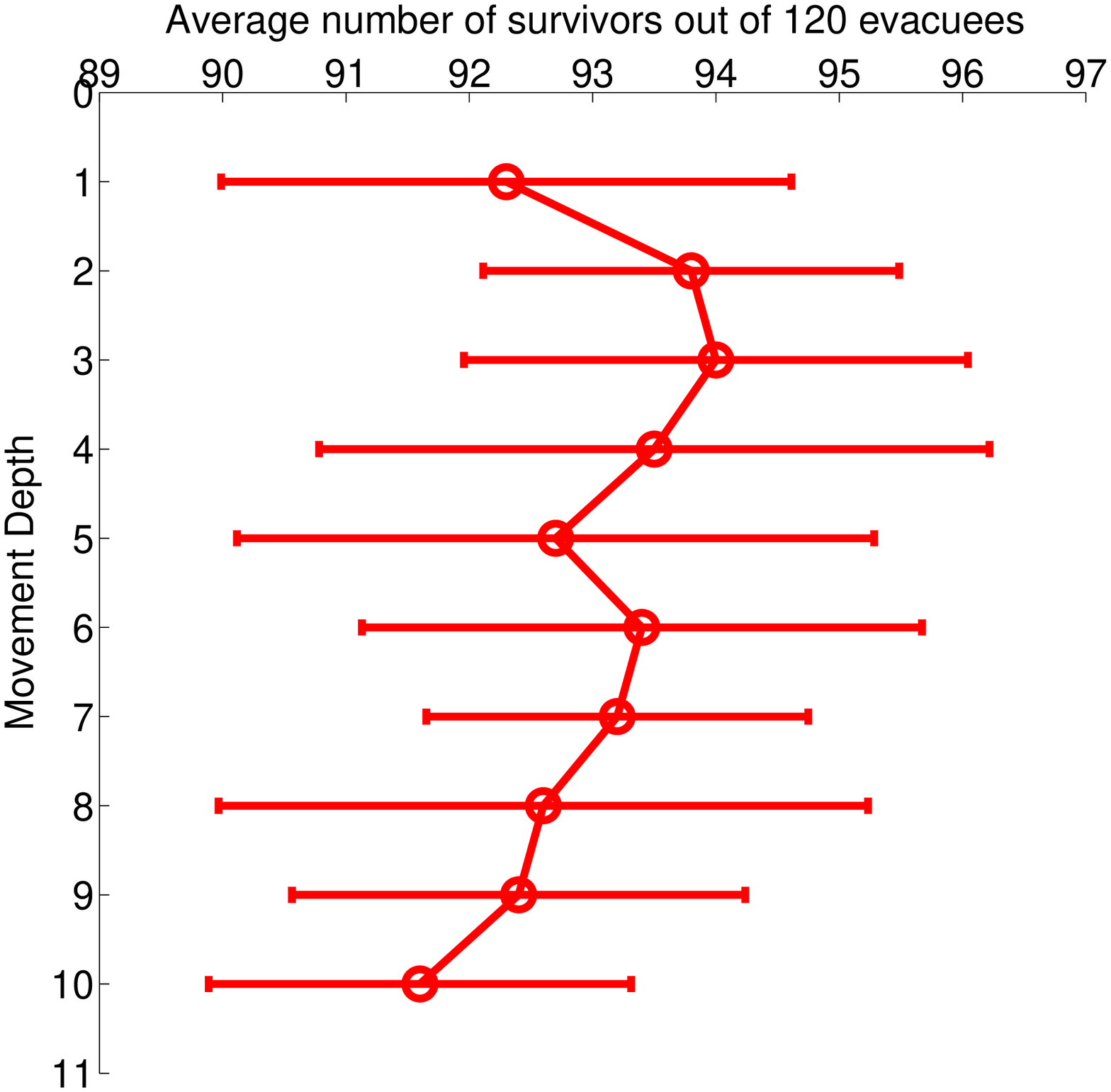}
\caption{Average number of survivors out of 120 evacuees with different movement depths. The error bars show the standard deviation of 10 experiments.}
\label{fig: oscillationshortesttime}
\end{figure}

Ten randomized iterations have been conducted for movement depth from 1 to 10 in a scenario with normal fire spreading rate and 120 evacuees. We use this scenario because the effect of oscillation is more obvious in high density environments due to the larger variation of congestion. Fig. \ref{fig: oscillationshortesttime} shows that the performance of the system reaches the peak mean value when the movement depth is 3 and then decreases gradually with the increase of the movement depth. This is because when movement depth is small (1 or 2), though it uses the up-to-date suggestion, some evacuees may be temporarily trapped by the puzzled suggestion. Meanwhile, CPN evaluates the QoS metric of a whole path other than a single edge. In other words, the evacuees should finish the whole path to satisfy their QoS needs if the condition does not change. If they continuously change paths, they will not follow the optimal path but a collection of the first edge of optimal paths. If movement depth is too large, evacuees will use the outdated decisions and therefore increase the possibility of guiding evacuees near hazard and backtracking. Hence, we set the movement depth to 3 in our experiments.

\section{Simulation and Experiments} \label{Sim}

The Distributed Building Evacuation Simulator (DBES) \cite{DBES} which is an agent based discrete event simulator where each entity is modelled as a process,
and the physical world is represented by a set of ``Points of Interest'' (PoI) which are significant locations such as doorways and corridors, and links which are physical paths.

\begin{figure}[!t]
\centering
\includegraphics[width=0.5\textwidth]{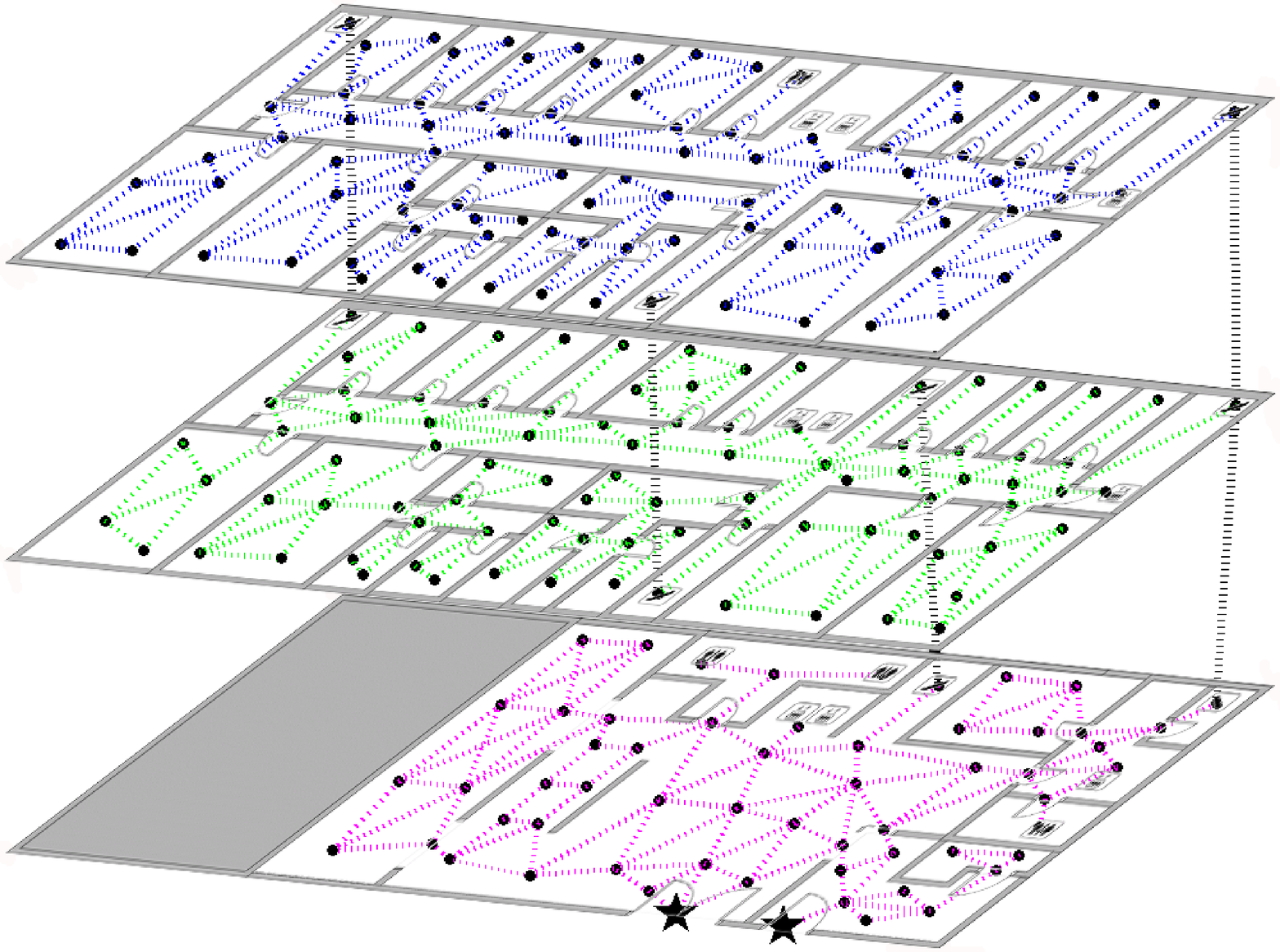}
\caption{The graph based layout of the building. The black stars represent the exits on the ground floor.}
\label{fig: graph3D}
\end{figure}

The environment is the lower floors of Imperial College's EEE building shown in Fig. \ref{fig: graph3D}. SNs and DNs are deployed in proximity to PoI. SNs collect hazard information and DNs execute the decision support algorithms. Real time decisions can be transformed to evacuees with visual indicators that are installed at POI or portable devices carried by the civilians. To evaluate the performance and adaptability of decision algorithms, we assume a fire outbreak near the eastern staircase on the ground floor that blocks a main channel. Initially, evacuees are randomly scattered in the building and that they hear or see fire alarms and evacuate immediately.
Three routing algorithms are employed: Dijkstra's shortest path algorithm, CPN based algorithm with distance metric and CPN with time metric. In first scenario, when the fire spreads, global hazard information will be synchronized in the network and Dijkstra's shortest path algorithm will be executed at each decision node to obtain the up-to-date shortest path. The second and the third scenario are similar to the first one but using the CPN based algorithm to collect information and make decisions.

\section{Results and Discussion} \label{Discuss}

The experiments are conducted on scenarios with 30, 60, 90, 120 evacuees, respectively. The Dijkstra's
shortest path algorithm and original distance-oriented goal function based experiments are performed for comparison purpose. To exclude the interferences of other factors, the evacuees in the simulation are homogeneous with identical velocity and initial health value.

\begin{figure}[htb]
\centering
\includegraphics[width=0.48\textwidth,height=6cm]{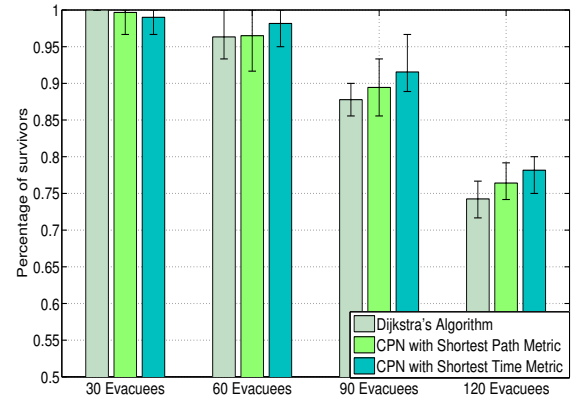}
\caption{Percentage of survivors for each scenario. The results are the average of 10
randomized simulation runs, and error bars shows the min/max result in any of the 10
simulation runs.}
\label{fig: timeoriented}
\end{figure}

The results of ten randomised experiments are presented in Fig \ref{fig: timeoriented}. The error bars show the highest and lowest number of survivors in the ten simulations. In low occupancy rate (30 evacuees), both CPN with time metric and CPN with distance metric reach the performance of Dijkstra's algorithm which can be considered as an optimal algorithm because it has full knowledge of the environment and the level of congestion is low. As the population density increases, congestion occurs more and more frequently and CPN with time metric becomes the best algorithm because it takes the potential congestion into account and searches the path with shortest evacuation time. While Dijkstra's algorithm and CPN with distance metric insist on one main channel to the exits, CPN with time metric can distribute evacuees to several channels and effectively ease congestion, as can be seen from Table \ref{table: numberofCongestion}.

\begin{figure}[ht!]
\centering
\includegraphics[width=0.48\textwidth,height=6cm]{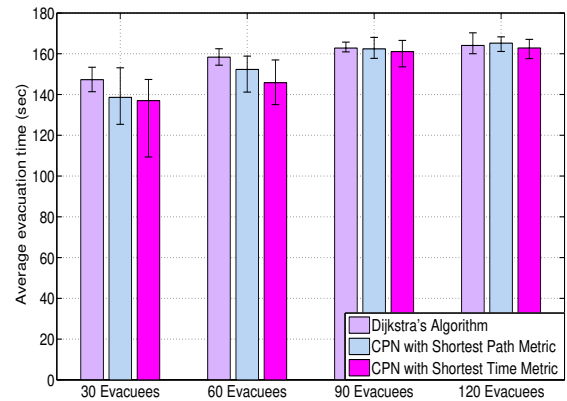}
\caption{Average evacuation times of ten iterations in arbitrary time units. The error bars represent the min/max values found in the ten simulations.}
\label{fig: averageevacuationtime}
\end{figure}

Fig. \ref{fig: averageevacuationtime} shows the average evacuation time in the above scenarios and the error bars represent the maximum and minimum egress time of ten experiments. Results confirm that CPN with time metric achieve the average shortest evacuation time among the three metrics. Furthermore, both CPN based algorithms can accomplish an evacuation process faster than Dijkstra's algorithm in low occupancy rates (30 and 60 evacuees) while gaining similar number of survivors. However, the three metrics achieve comparable results in high population densities (90 and 120 evacuees), with a slightly advantage to CPN with time metric. This is mainly caused by congestion formed at the bottlenecks of the network. A large amount of evacuees have to queue for a long time at the central staircase between the ground floor and the first floor before they can reach the ground floor. The maximum queue length at this staircase can attain 25 for Dijkstra's algorithm and 20 for CPN with distance metric. Interestingly, although the value for CPN with time metric is 16, alternate congestion will form near exits as the egresses become new bottlenecks. It proves that in high density environment, the bottlenecks in the system can significantly affect the average evacuation time. In other words, though the CPN with time metric can accelerate the process for civilians to arrive in the vicinity of exits, it can not significantly reduce the average evacuation time under the equivalent departure rate of the exits.

\begin{figure}[ht!]
\centering
\includegraphics[width=0.48\textwidth,height=5cm]{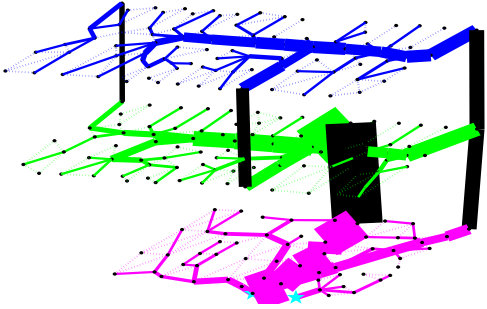}
\caption{Dijkstra's algorithm routing}
\label{fig: routedijkstra}
\end{figure}

\begin{figure}[ht!]
\centering
\includegraphics[width=0.48\textwidth,height=5cm]{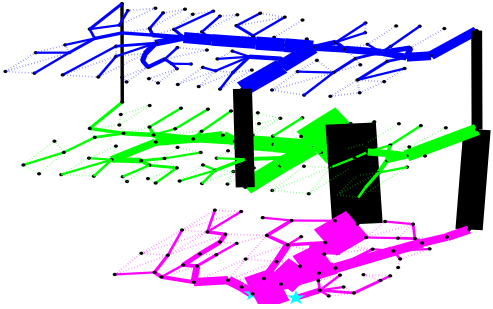}
\caption{Shortest-path metric routing}
\label{fig: routeshortestpath1}
\end{figure}

\begin{figure}[ht!]
\centering
\includegraphics[width=0.48\textwidth,height=5cm]{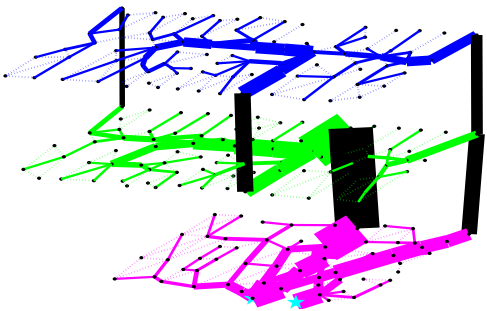}
\caption{Shortest-time metric routing}
\label{fig: routeshortesttime1}
\end{figure}

Fig. \ref{fig: routedijkstra} - \ref{fig: routeshortesttime1} show the number of visits of each edge in a scenario with 120 evacuees. The thickness of solid line is proportional to the number of visits. The edges that have not been traversed are shown as dotted lines. Fig. \ref{fig: routedijkstra} and Fig. \ref{fig: routeshortestpath1} which use the shortest path routing metric shows that the shortest egress path on the ground floor are overused by evacuees. Most evacuees select the identical main channel to egress and cause continuous congestion. In both Dijkstra's algorithm and CPN with distance metric, 96\% of the evacuees use the main exit in the middle and only 4\% of the evacuees escape from the alternate exit. Fig. \ref{fig: routeshortesttime1} which employs the shortest time routing metric shows that several channels are formed on the ground floor to transmit evacuees. This phenomenon does not appear on the other floors because the corridors on the first and second storey are narrow. Meanwhile, the distribution of evacuees are more balanced between the two exits, with 79\% of the evacuees depart from the main exits and 21\% of the civilians choose the other exit. Therefore, the probability of congestion is reduced and the latency in the evacuation process is decreased.

\begin{table}
    \begin{center}
        \begin{tabular}{| r | c | c | c |}
            \hline
            Number of Evacuees & Dijkstra & CPN with SP & CPN with ST \\ \hline
            30 ~~~~~~~~ & 97.8 & 97.6 & 80.8 \\ \hline
            60 ~~~~~~~~ & 382.8 & 363.5 & 332.7 \\ \hline
            90 ~~~~~~~~ & 695.9 & 676.8 & 627.7 \\ \hline
            120 ~~~~~~~ & 943.5 & 935.7 & 850.3 \\ \hline
            \hline
        \end{tabular}
        \caption{Average number of congestion appears in each scenario for different path finding algorithms. Congestion is assumed to happen when an evacuee reaches a node with a non-zero queue length. ``CPN with SP'' presents CPN based algorithm with distance metric and ``CPN with ST'' depicts CPN based algorithm with time metric.}
        \label{table: numberofCongestion}
    \end{center}
\end{table}

\bibliographystyle{IEEEtran}
\bibliography{esa}

\end{document}